\documentclass[twocolumn]{aastex631} 
\usepackage{amsmath}
\usepackage{graphicx}
\usepackage{natbib}
\usepackage{gensymb}
\usepackage{multirow}
\usepackage{color, soul}
\usepackage{array}
\usepackage{soul}

\renewcommand{\arraystretch}{1.5}
\newcommand{\W}{\,{\rm W}}
\newcommand{\m}{\,{\rm m}}
\newcommand{\erg}{\,{\rm erg}}

\shortauthors{Sinha et al.}

\begin{document}

\title{A Comparative Analysis of Machine-learning Models for Solar Flare Forecasting: Identifying High-performing Active Region Flare Indicators}

\author[0000-0001-7229-5192]{Suvadip Sinha}
\affiliation{Center of Excellence in Space Sciences India, Indian Institute of Science Education and Research Kolkata, Mohanpur 741246, West Bengal, India}

\author[0000-0001-8470-7289]{Om Gupta}
\affiliation{Center of Excellence in Space Sciences India, Indian Institute of Science Education and Research Kolkata, Mohanpur 741246, West Bengal, India}
\affiliation{Department of Physical Sciences, Indian Institute of Science Education and Research Kolkata, Mohanpur 741246, West Bengal, India}

\author[0000-0002-9911-2285]{Vishal Singh}
\affiliation{Center of Excellence in Space Sciences India, Indian Institute of Science Education and Research Kolkata, Mohanpur 741246, West Bengal, India}
\affiliation{Department of Physical Sciences, Indian Institute of Science Education and Research Kolkata, Mohanpur 741246, West Bengal, India}

\author[0000-0003-0666-7650]{B. Lekshmi}
\affiliation{Center of Excellence in Space Sciences India, Indian Institute of Science Education and Research Kolkata, Mohanpur 741246, West Bengal, India}
\affiliation{Max Planck Institute for Solar System Research, Göttingen, Germany 37077}

\author[0000-0001-5205-2302]{Dibyendu Nandy$^*$}
\correspondingauthor{Dibyendu Nandy}
\email{dnandi@iiserkol.ac.in}
\affiliation{Center of Excellence in Space Sciences India, Indian Institute of Science Education and Research Kolkata, Mohanpur 741246, West Bengal, India}
\affiliation{Department of Physical Sciences, Indian Institute of Science Education and Research Kolkata, Mohanpur 741246, West Bengal, India}

\author[0000-0003-4861-8152]{Dhrubaditya Mitra}
\affiliation{Nordita, KTH Royal Institute of Technology and Stockholm University,
Hannes Alfv\'ens v\"ag 12, SE-10691 Stockholm, Sweden}

\author[0000-0003-2638-6047]{Saikat Chatterjee}
\affiliation{KTH Royal Institute of
Technology, Stockholm}

\author[0000-0001-5220-1881]{Sourangshu Bhattacharya} 
\affiliation{Indian Institute of Technology, Kharagpur, India}

\author[0000-0002-7817-9851]{Saptarshi Chatterjee} 
\affiliation{Indian Institute of Technology, Kharagpur, India}

\author[0000-0002-0452-5838]{Nandita Srivastava}
\affiliation{Udaipur Solar Observatory, Physical Research Laboratory, P.O. Box 198, Badi Road, Udaipur 313001,
India}
\affiliation{Center of Excellence in Space Sciences India, Indian Institute of Science Education and Research Kolkata, Mohanpur 741246, West Bengal, India}

\author[0000-0002-7304-021X]{Axel Brandenburg}
\affiliation{Nordita, KTH Royal Institute of Technology and Stockholm University,
Hannes Alfv\'ens v\"ag 12, SE-10691 Stockholm, Sweden}
\affiliation{The Oskar Klein Centre, Department of Astronomy,
Stockholm University, AlbaNova, SE-10691 Stockholm, Sweden}

\author[0000-0002-6302-438X]{Sanchita Pal}
\affiliation{Department of Physics, University of Helsinki, P.O. Box 64, FI-00014 Helsinki, Finland}

\begin{abstract}

Solar flares create adverse space weather impacting space and Earth-based technologies. However, the difficulty of forecasting flares, and by extension severe space weather, is accentuated by the lack of any unique flare trigger or a single physical pathway. Studies indicate that multiple physical properties contribute to active region flare potential, compounding the challenge. Recent developments in machine learning (ML) have enabled analysis of higher-dimensional data leading to increasingly better flare forecasting techniques. However, consensus on high-performing flare predictors remains elusive. In the most comprehensive study to date, we conduct a comparative analysis of four popular ML techniques (k-nearest neighbor, logistic regression, random forest classifier, and support vector machine) by training these on magnetic parameters obtained from the Helioseismic and Magnetic Imager (HMI) on board the Solar Dynamics Observatory (SDO) for the entirety of solar cycle 24. We demonstrate that the logistic regression and support vector machine algorithms perform extremely well in forecasting active region flaring potential. The logistic regression  algorithm returns the highest true skill score of $0.967 \pm 0.018$, possibly the highest classification performance achieved with any strictly parametric study. From a comparative assessment, we establish that the magnetic properties like total current helicity, total vertical current density, total unsigned flux, R\_VALUE, and total absolute twist are the top-performing flare indicators. We also introduce and analyze two new performance metrics, namely, severe and clear space weather indicators. Our analysis constrains the most successful ML algorithms and identifies physical parameters that contribute most to active region flare productivity.    
 
\end{abstract}

\section{Introduction}
\label{sec:intro}
Solar flares are sudden bursts of electromagnetic radiation from the solar
atmosphere, mainly in the extreme ultraviolet and X-ray regimes.
They are classified into different categories based on the peak X-ray flux
recorded in the 1--$8\,${\AA} band by the Geostationary Operational Environmental Satellite (GOES).  The X-class flares are the most powerful with peak fluxes $\ge 10^{-4}\W\m^{-2}$, followed by M-class flares with peak fluxes $\ge 10^{-5}\W\m^{-2}$. These classes of flares strongly influence the near-Earth space weather and present a bigger potential hazard to human space endeavors than flares with lower peak intensities, which, in decreasing order of intensity, belong to the C, B, and A classes.

From previous studies we know that solar flares originate in active region (AR) structures, where the magnetic flux system  becomes energized due to rapid flux emergence, instability, or topological changes of the magnetic configuration via reconnection processes \citep{Forbes2000, PriestForbes2002, Schrijver2007, Leka2003a, Leka2003b, nandy2003, hahn2005, Jing2006}. A solar AR with a potential or near-potential magnetic field builds up magnetic nonpotential energy (or free magnetic energy) upon being sheared and twisted.
A fraction of this free energy is dissipated during a flare event (e.g., \citealt{Schrijver2008}), and a typical large solar flare can release large quantities of energy of the order of $10^{32}$--$10^{33}\erg$. Simultaneously, solar energetic particles are also released into the solar wind. Solar flares are often accompanied by Coronal Mass Ejections (CMEs), which pose serious threats if directed towards the Earth. Earlier studies have shown that the magnetic characteristics of ARs \citep{Yeates_2010, Pal2018, 2017ApJ...851..123P} and filaments determine their propensity to flare and produce associated CMEs \citep{Sinha2019}. 

Solar flares (and CMEs) induce extreme space weather conditions that have the potential to harm satellites and impact communication and navigation sectors.
The most energetic solar flares can cause electric power grid failures and radio communication blackouts, impact astronaut health, and expose air passengers to harmful doses of radiation \citep{Hapgood2011, Schrijver2015a, Schrijver2015b, Eastwood2017}. Proactive measures to mitigate the physical and economic impact of space weather are therefore much sought after, of which early-warning systems are of foremost interest. While physical model based studies have demonstrated a strong potential for success in recent times towards predicting long-term solar activity variations over decadal timescales \citep{Bhowmik2018, Nandy2021, 2021PEPS....8...40N}, physical model based assessment of AR flaring probability remains elusive.

The creation of knowledge toward predicting solar flares had been initiated with statistical approaches applied to observational data well before machine-learning (ML) techniques found favor. In a set of pioneering studies with vector magnetogram data \citet{Leka2003b} and \citet{Barnes2007} conducted a multiparametric statistical study to distinguish between flaring and flare-quiet ARs based on discriminant analysis.

One of the early applications of ML in solar physics was the automatic real-time detection of solar flares from H$\alpha$ images (e.g., \citealt{Borda2002, Qu2003}). Very soon, efforts were directed toward the 
forecast of solar flares. A number of ML methods were trained on sunspot-associated data to forecast solar flares \citep{Li2007, Qahwaji2007, Benvenuto2018, Cinto2020}. \citet{Colak2009} used neural networks to make multiclass forecasts based on sunspot area and McIntosh classification data.
Line-of-sight full-disk magnetogram data from the Solar and Heliospheric Observatory's (SOHO) Michelson Doppler Imager (MDI) presented the next opportunity in the development of solar flare forecasting methods and several advances were made by using features calculated from them. Decision tree classifiers, learning vector quantization, ordinal logistic regression, support vector machine (SVM), and AdaBoost methods were experimented with by \citet{Yu2009}, \citet{Song2009}, \citet{Yuan2010}, \citet{Huang2010} and \citet{Lan2012}. \citet{Ahmed2013} applied a cascade correlation neural network and used feature evaluation algorithms to remove redundant features and showed that a smaller set of parameters yielded comparable results to the entire set.
\citet{Huang2018} combined the line-of-sight magnetograms from MDI and the Solar Dynamics Observatory's (SDO) Helioseismic and Magnetic Imager (HMI) to create an extensive data set and evaluated the performance of a Convolutional Neural Network (CNN) on these data.

Following the launch of SDO in 2010, its HMI instrument \citep{2012Scherrer}
started providing one of the most advanced unhindered full-disk vector magnetograms. To facilitate AR-based event forecasting, the Spaceweather HMI Active Region Patch (SHARP) data product~\citep{2014Bobra} provides cutouts of automatically tracked magnetic flux concentrations on the solar disk. Using SHARP data, \citet{Bobra2015} implemented an SVM algorithm to distinguish between ARs producing an M- or X-class flare (in the next 24 hr) and those not producing any flare or low-intensity flares.
\citet{Bobra2015} presented a significant improvement in the performance of AR-parameter based ML algorithms, primarily due to the availability of continuous, high-quality HMI vector magnetogram data for deriving input magnetic features.

These recent advances piqued the interest of both the solar physics and computer science communities heralding a close interdisciplinary collaboration in solar flare forecasting. \citet{Liu2017} attempted a multiclass classification using random forest; \citet{Nishizuka2017} and \citet{Florios2018} compared various ML algorithms which included SVM, multi-layer perceptrons, random forest and k-nearest neighbors (KNN) algorithm, \citet{Nishizuka2018,Nishizuka2020} trained a deep neural network for binary classification, and \citet{Campi2019} used hybrid LASSO and random forest algorithms on features derived during the FLARECAST project. In a recent study, \cite{RIBEIRO2021100468} used LightGBM for flare forecasting and showed a nice comparison with existing ML models.  
Classification using KNN was attempted by \citet{Hamdi2017} for univariate time series and by \citet{Boubrahimi2018} for multivariate time series. Decision trees were used by \citet{Ma2017} for multivariate time series. \citet{Liu2019} implemented time series classification by training long short-term memory (LSTM) neural networks on SHARP features and flare history parameters. A similar approach was followed by \citet{Jiao2020} who built classification models on an LSTM regressor. \citet{Chen2019} compared LSTM models trained on SHARP parameters and autoencoder-derived features.  Using wavelet analysis and features derived from SDO HMI magnetograms, support vector regression was applied to the forecasting of the X-ray flux by \citet{Muranushi2015} and \citet{Boucheron2015}, while \citet{AlGhraibah2015} attempted classification using relevance vector machines. Zernike moments calculated from images were also used for binary classification with SVM \citep{Raboonik2017, Alipour2019}.
Strong-field high-gradient polarity inversion line (PIL) features derived from SHARP images were used by \citet{Sadykov2017} for classification comparing SVM and a graphical method, while \citet{Wang2019} used SHARP parameters weighed with a PIL mask to improve individual parameter performance on a random forest classifier. \citet{Dattaraj2019} and \citet{Hazra2020} studied the time evolution of various magnetic parameters and the correlations between them. They trained and tested LR, SVM, gradient boost, random forest, multilayer perceptron, KNN and a naïve Bayes classifier on SHARP feature data with good performance. 

With rapid developments in the field of ML and image processing, it became possible to process images directly using CNNs. \citet{Jonas2018} used vector magnetic field data from HMI as well as multiwavelength image data of the chromosphere, transition region, and corona to train a single-layer CNN, and obtained results comparable to those of \citet{Bobra2015}. \citet{Zheng2019}, \citet{Li2020} and \citet{Bhattacharjee2020} used line-of-sight magnetograms to train deep CNN models.

The underlying nonunique and nondeterministic nature of the triggering mechanisms without well-defined parametric thresholds makes flare forecasting a challenging task making the problem suitable for multiparametric statistical approaches and computational ML algorithms applied to large databases.
Attempts to supplement vector magnetogram data with extreme ultraviolet images have not yielded significant improvement. On comparing CNN models trained with and without multi-wavelength image data from the Atmospheric Imaging Assembly (AIA) on board the SDO, \citet{Jonas2018} found that the best-performing model
was the one not provided with AIA data as input. Similarly, the implementation of CNNs has to be developed further for application in flare
forecasting. \citet{Bhattacharjee2020} found that the CNN output has spurious dependencies on the magnetogram dimensions.

We limit our comparative analysis to well-studied and successful ML algorithms (limited to parametric approaches alone for efficiency) to determine their relative performance. This is achieved by applying these algorithms to the largest, single-instrument database suitable for this purpose, i.e., the HMI vector magnetic field observations.

Over the last two decades a wide range of ML algorithms have been applied to forecast solar flares. The input data to such algorithms are, most commonly, several AR magnetic parameters derived from magnetograms, magnetogram images, and time series data of magnetic parameters. These works have been able to achieve reasonably accurate 
forecasts of whether an AR is going to flare or not and, if it does, in which class the flare lies in. Furthermore, these works have attempted to extract which magnetic parameters are best correlated with flaring. In general the obtained results are independent of the algorithms used -- unsigned current density, unsigned flux, and current helicity have come up as key parameters in most of the previous studies. However, their relative ranking in terms of which contribute most to the flare potential has not been rigorously explored. 

In this paper, we compare several ML algorithms to find out which offers the best flare forecasting capability. \citet{Bobra2015} and subsequent studies have shown that the magnetic twist parameter in the SHARP database does not perform well for machine classification whereas earlier physics-based works suggest that twist is a flare-relevant parameter \citep{Linton1996, nandy2003, hahn2005, Nandy2008Review}. We therefore include a global indicator of magnetic twist in our analysis. Furthermore we introduce two new performance metrics, the severe space weather (SSW) and clear space weather (CSW) indicators, to distinguish between these two equally important conditions. Our analysis, detailed in the followings sections, is based on the highest number of unique ARs to date used in training ML algorithms, from the beginning of the SDO era to 2020 December, covering the entire solar cycle 24.

\section{Data Selection}

Based on their intensities, flares are categorized into five classes: A, B, C, M, and X in the ascending order of intensity. In this study, all the AR information is collected from the $hmi.sharp\_cea\_720s$ series \citep{2014Bobra} but for a longer time of observation. We build our data set considering all the ARs that have appeared on the Sun starting from 2010 May to 2020 December. We divide the ARs into two groups -- the positive or flaring class and the negative or nonflaring class. The positive class is defined such that it consists of ARs that have produced at least one M- or X-class flare in their lifetime. In contrast, the negative class is formed by those ARs that produce only low-intensity flares ($\leq$ B-class) or do not flare at all.

\begin{figure}
    \centering
    \includegraphics[width=\linewidth]{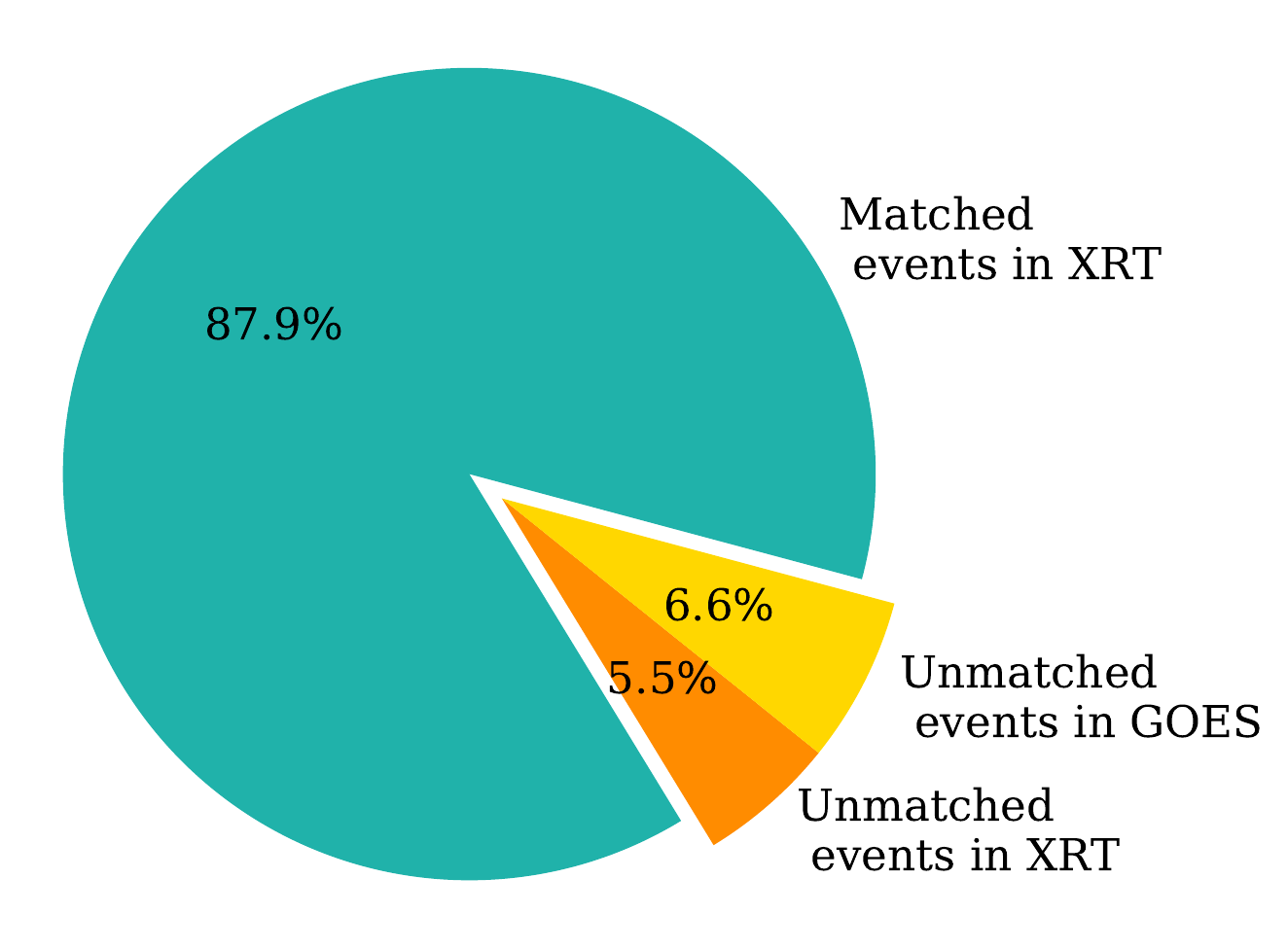}
    \caption{Distribution of matched and unmatched events between XRT and GOES flare catalogs.}
    \label{fig: pie-chart}
\end{figure}

We use the XRT Flare Catalog based on the Hinode Flare Catalogue \citep{Watanabe2012}, and the GOES flare catalog to collect information on flare events of the past 10 years, such as flare timing, flare intensities, associated ARs (NOAA number), and their positions on the solar disk. GOES flare events are collected from the sunpy.instr.goes module of the Python SunPy library \citep{sunpy}.
Our data set covers the entire solar cycle 24 starting from 2010 May to 2020 December. We find that not all events in the XRT catalog match the GOES event list because of slight differences in flare peak times.
For each event in the XRT catalog we search for a similar event in the GOES catalog within a time window of 4 minutes centered at the flare peak time of the XRT event.
If a similar flare event is found in the GOES catalog with the same flare class and NOAA number, we call it a match.
The pie chart in Figure \ref{fig: pie-chart} depicts the number of matched and unmatched events in the two catalogs. Following this, a total of 668 matched and 80 unmatched events in the XRT catalog are obtained with flare intensity $\geq$ M-class.
Manual inspection of these 80 unmatched events with more lenient conditions, for example, by allowing flare peak time differences up to 15 minutes, results in a further reduction of the number of unmatched events by 57. The negative data set is prepared by excluding flare-associated ARs from all recorded SHARP regions during our observational time domain.

All the magnetic parameters representing the flaring AR are either collected from the SHARP header keywords or calculated from the vector magnetic field data, 24 hr before the flare peak time. For the negative/nonflaring class, we choose the magnetogram observation at the central time snap of their entire lifespan on the visible disk. In addition, if an AR's position vector from the Sun center makes an angle greater than $70^\circ$ with the line of sight, we discard that region from our analysis, which is a standard method to avoid high projection effects.
We implement this $70^\circ$ positional filter in the last step of our data preparation process so that it only restricts our domain of analysis, not the domain of our observation. In other words, this ensures that any AR producing an M/X-class flare outside this $70^\circ$ angular region is not included in our negative class.

Following all these selection criteria, our final positive class contains 503 flaring events and the negative class consists of 3358 nonflaring events. Note that in our positive class, recurrent flare events are treated as separate events with different entries.
In contrast, each nonflaring SHARP region has a single entry in the negative class.

\section{Methods and analysis}

\begin{figure}
    \centering
    \includegraphics[width=\linewidth]{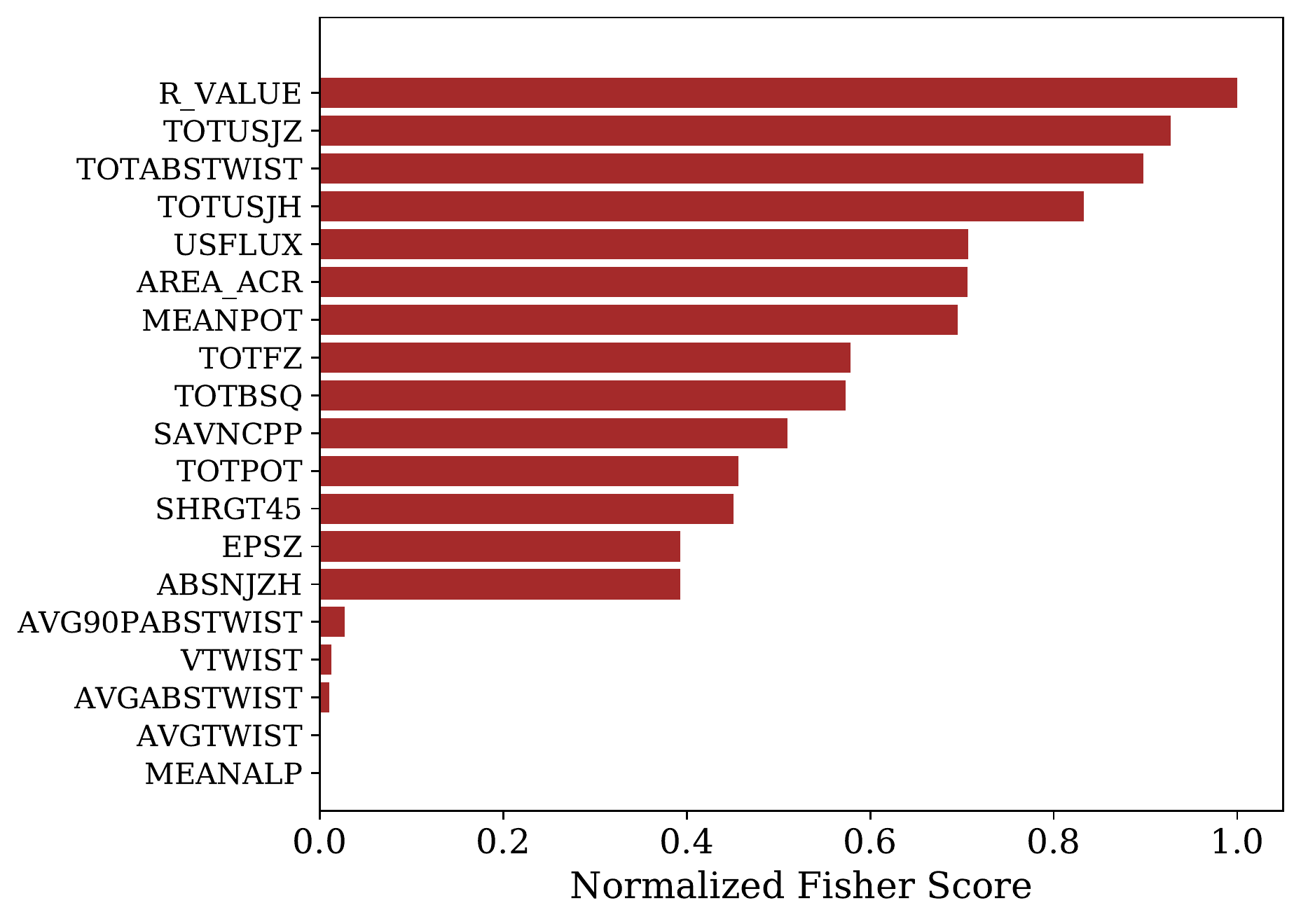}
    \caption{Normalized F-score ranking of the magnetic field parameters.}
    \label{fig:F-score}
\end{figure}

Preparing the input data and properly training the model are the most crucial steps in working with ML algorithms \citep{ahmadzadeh2021}. 
SHARP data comes with various magnetic parameters, calculated from the vector magnetic field maps of the ARs.
Previous studies showed the importance of these derived parameters in characterizing AR properties and complexities \citep{1984Hagyard,a2003Leka,b2003Leka,Georgoulis2007,2007LaBonte,2012Moore}. \cite{Bobra2015} identified 13 such parameters that they found to be most useful in describing the flare potential. However, they estimated the magnetic twist using the parameter called MEANALP, whose poor performance led them to exclude any contribution of the magnetic twist from their classification. But it has been shown that a high twist in the magnetic flux rope can store nonpotential magnetic energy and often leads to eruption of the flux rope via the kink instability \citep{Nandy2008}.
Motivated by these physical arguments, we calculate six new magnetic parameters related to AR twist, such as TOTABSTWIST, AVG90PABSTWIST, VTWIST, AVGABSTWIST, AVGTWIST, and MEANALP (see Table 1 for descriptions). We incorporate these 6 twist-related features along with those 13 parameters used by \cite{Bobra2015}. Here we assume the force-free field approximation for the estimation of magnetic twist -- also known as the alpha parameter.
The vertical twist parameter $\alpha_z$ at each pixel of an AR magnetogram is given as

\begin{equation}
\alpha_z=\mu_0  \left(\frac{J_z}{B_z}\right ),
\end{equation}
where $J_z$ and $B_z$ are the $z$-components of the current density and magnetic field, respectively,
and $\mu_0$ is the permeability of free space.
All these 19 features and their descriptions are listed in Table \ref{tab:sharp_param}.
We perform univariate feature selection analysis with the ANOVA
Fisher statistics (F-statistics) using the python scikit-learn library
to finalize our set of input magnetic features by eliminating those that are not very useful for this classification.
The obtained F-score (see \cite{Bobra2015} for the calculation of F-score) for each feature is represented in Figure~\ref{fig:F-score}.
It is quite surprising to note that all the magnetic twist related parameters (including MEANALP) are insignificant according to the F-statistics except for the TOTABSTWIST, which ranks third in the list. This indicates that flare potential is more closely coupled to gross/extrinsic quantities than to magnetic properties at individual pixels.
We exclude the last five features having the lowest normalized F-scores in Figure \ref{fig:F-score}, i.e., AVG90PABSTWIST, VTWIST, AVGABSTWIST, AVGTWIST, and MEANALP (descriptions in Table \ref{tab:sharp_param}). All further analyses are done with the remaining 14 features, which contain 10 SHARP keyword parameters and 4 derived parameters including 1 newly introduced parameter TOTABSTWIST.

\begin{table*}[]
    \centering
    \begin{tabular}{|c|c|c|c|}
        \hline
        S. No. & Keyword & Description & Normalized F-score \\
        \hline
        \hline
        1 & R\_VALUE & Sum of unsigned flux near PIL & 1.000 \\
        2 & TOTUSJZ & Total unsigned vertical current & 0.927\\
        3 & TOTABSTWIST* & Total absolute twist calculated over strong-field ($\left |B\right | \geq 300\,$G) regions & 0.898 \\
        4 & TOTUSJH & Total unsigned current helicity & 0.833 \\
        
        5 & USFLUX & Total unsigned flux & 0.707 \\
        6 & AREA\_ACR & Area of strong-field pixels in the AR & 0.706 \\
        7 & MEANPOT & Mean photospheric magnetic free energy & 0.695 \\
        8 & TOTFZ* & Sum of $z$-component of Lorentz force & 0.579 \\
        9 & TOTBSQ* & Total magnitude of Lorentz force & 0.573 \\
       10 & SAVNCPP & Sum of the modulus of the net current per magnetic polarity & 0.510 \\
       11 & TOTPOT & Total photospheric magnetic free energy density & 0.456\\
       12 & SHRGT45 & Fraction of area with shear $> 45 \degree$ & 0.451 \\
       13 & EPSZ* & Sum of z-component of normalized Lorentz force & 0.393 \\
       14 & ABSNJZH & Absolute value of the net current helicity & 0.393 \\
       15 & AVG90PABSTWIST* & Average absolute twist for pixels having twist more than 90th percentile value & 0.027 \\
       16 & VTWIST* & Standard deviation of twist within an AR & 0.013 \\
       17 & AVGABSTWIST* & Average absolute value of twist & 0.011 \\
       18 & AVGTWIST* & Average value of twist & 0.000 \\
       19 & MEANALP & Mean value of flux-weighted twist & 0.000 \\
        \hline
    \end{tabular}
    \caption{Details of AR parameters extracted from SHARP Data. Asterisk (*) denotes parameters that are not readily available in the SHARP header keywords and are calculated explicitly from the SHARP vector magnetic field data.}
    \label{tab:sharp_param}
\end{table*}

\begin{figure*}
    \centering
    \includegraphics[width=\linewidth]{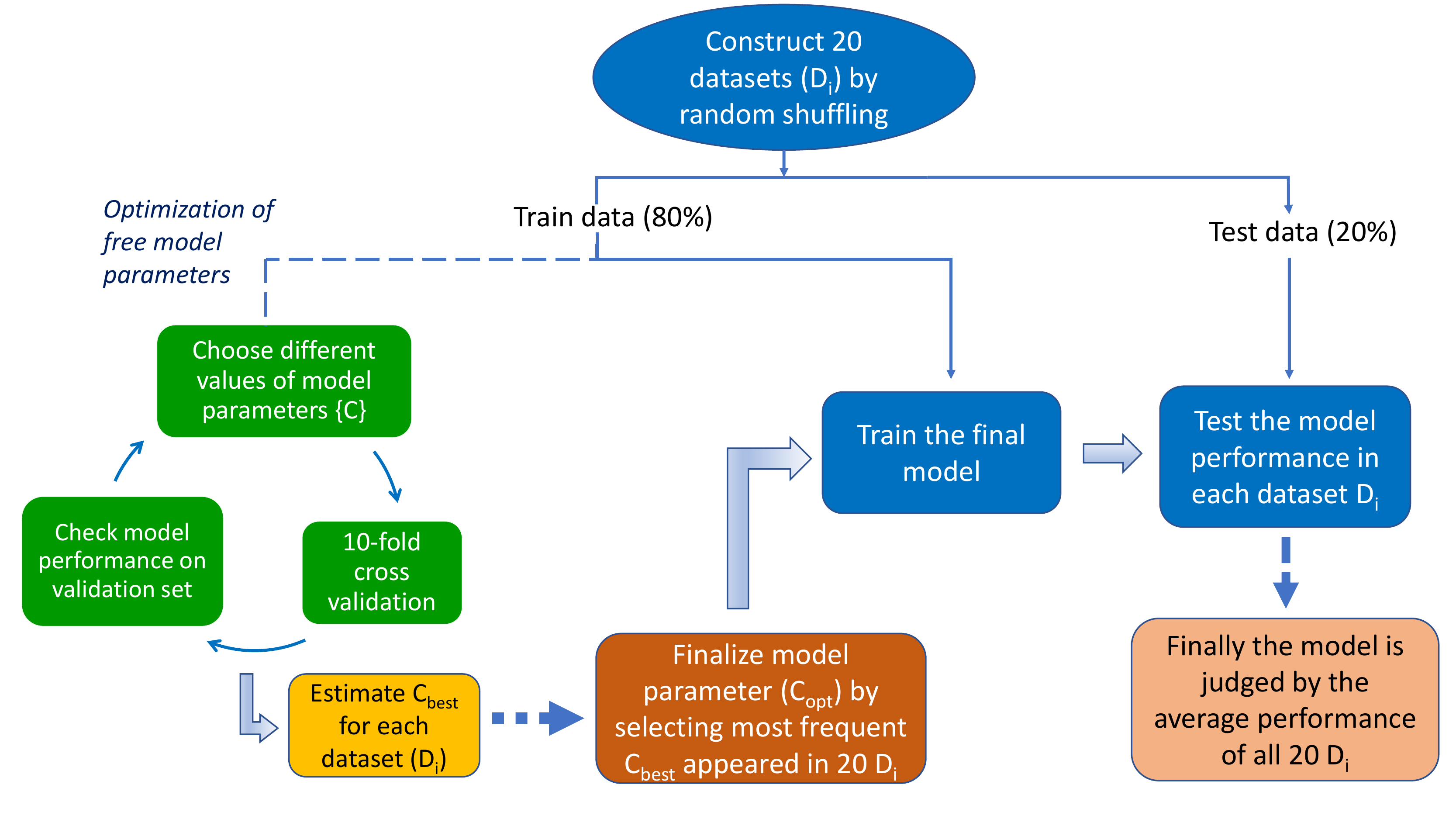}
    \caption{Schematic diagram of our method of analysis.}
    \label{fig:schematic}
\end{figure*}

Henceforth, each AR is represented by a single data point in 14-dimensional feature space except those ARs that have produced multiple M/X-class flares. The latter are accounted as separate events for each M/X-class flares. Our whole data set, consisting of 3861 events, is randomly divided into two groups, training and testing, with a population ratio of 4:1 respectively. We arrange the data such that the ratio of flaring to nonflaring events is the same for both the training set and the test set. We preprocess the data by normalizing it such that the processed data has zero mean and unit standard deviation. For this normalization we solely use the training data set and then apply the same population mean and standard deviation to normalize the test data set. To make our classification more robust and independent of any bias, we randomly shuffle our data set to make 20 similar but differently distributed representative pairs of training and testing data.
We denote each such pair with $D_{\rm i}$ where $i$
is a running index between 1 and 20.
Each ML algorithm is evaluated by its average performance over these 20 $D_{\rm i}$.
A schematic diagram of our analysis method is shown in Figure \ref{fig:schematic}. The following two sections describe how we quantify model performance and compare between the different ML algorithms.

\subsection{Performance metric}
Typically, the performance of an ML model is evaluated from the confusion matrix. It is a $2 \times 2$ matrix whose elements are the number of correctly forecasted positive-class events (TP), the number of correctly forecasted negative-class events (TN), the number of events falsely forecasted as positive-class (FP), and the number of events falsely forecasted as negative-class (FN). In general, there are various parameters that can be derived from the confusion matrix such as accuracy, recall, F1-score, etc., but their suitability depends on the particular problem.
Simple accuracy is defined as the ratio of the number of correct forecasts to the total number of forecasts. 
In our data set, the number of flaring ARs (positive-class) is much less than the number of nonflaring (negative-class) ARs, which means our data set is highly imbalanced.
Hence we cannot simply use the accuracy measure to evaluate
the models.
To deal with this problem we use the macro accuracy (MAC)\citep{PEREIRA2018359} and true skill score (TSS) \citep{Woodcock1976} values to evaluate our ML models. \cite{Bloomfield2012} showed that the TSS is unaffected by class imbalance and gives an unbiased result. The MAC and TSS are defined as

\begin{subequations}
\begin{align}
    &{\rm MAC} = \frac{1}{2} \left( \frac{{\rm TP}}{{\rm TP+FN}} + \frac{{\rm TN}}{{\rm FP+TN}} \right)\/, \\
    &{\rm TSS} = \frac{{\rm TP}}{{\rm TP+FN}} - \frac{{\rm FP}}{{\rm FP+TN}}\/.
\end{align}
\end{subequations}

The MAC is the average of the accuracy
of each individual class; hence its value lies between 0 and 1. On the other hand TSS has two components: the first is the positive-class accuracy and the second is the probability of false forecasts for the negative-class. Using the TSS, we penalize the model's performance commensurately by subtracting the false-alarm ratio from the positive-class accuracy. This shows the usefulness of the TSS in the present problem as we are more interested in correctly predicting flaring ARs with a minimal number of false detections. The value of the TSS ranges from $+1$ to $-1$ and we optimize our models to maximize the TSS. 

Depending on the end-user application our priorities for detecting a specific class can change. For example, one may wish to identify only those regions that have the potential to flare with a high degree of confidence without worrying about missclassifying a nonflaring region as a flaring region. This motivates a new performance indicator, which we term the SSW metric, defined as
\begin{equation}
{\rm SSW} = \frac{{\rm TP-FN}}{{\rm TP+FN}}.
\end{equation}
Conversely, one may wish to focus on identifying nonflaring regions only with a high degree of confidence. For this we define another parameter called the CSW metric:

\begin{equation}
{\rm CSW} = \frac{{\rm TN-FP}}{{\rm TN+FP}}.
\end{equation}

It is important to note that SSW is only linked to flaring-class events. It indicates the correct identification ratio combined with a penalty for misidentification within the flaring class. Similarly, the CSW deals with the nonflaring class only. The value of these two metrics lies between $-1$ and $+1$, where $+1$ indicates perfect identification of all the events within a specific class, whereas $-1$ indicates the scenario where all events are misclassified. A metric score of 0 denotes the scenario where half of the events in a specific class are correctly identified and the other half are wrongly classified implying no useful classification capability.

Moreover the average value of these two performance metrics returns the TSS and can be demonstrated as follows:

\begin{flalign*}
    {\rm SSW+CSW} = \frac{\rm TP-FN}{\rm TP+FN} + \frac{\rm TN-FP}{\rm TN+FP}&&
\end{flalign*}
\begin{flalign*}
    = \frac{\rm TP}{\rm TP+FN} - \frac{\rm FP}{\rm TN+FP} + \frac{\rm TN}{\rm TN+FP} - \frac{\rm FN}{\rm TP+FN}
\end{flalign*}
\begin{flalign*}
    = \frac{\rm TP}{\rm TP+FN} - \frac{\rm FP}{\rm TN+FP} + \left ( \frac{\rm TN}{\rm TN+FP}-1 \right ) &&\\ + \left(1 - \frac{\rm FN}{\rm TP+FN}\right)
\end{flalign*}
\begin{flalign*}
    = 2 {\rm TSS} \implies \frac{{\rm SSW + CSW}}{{\rm 2}} = {\rm TSS}. &&
\end{flalign*}

\subsection{Cross-validation}

One of the most important aspects of any ML algorithm is the optimization of its hyperparameters, to achieve best fit on the data set. If a classifier performs too well on the training data set, it might fail to capture the overall picture and can badly perform on the test data set –- also known as overfitting.
The optimization of hyperparameters is done by employing a grid Search algorithm for finding the optimal hyperparameters $C_{\rm opt}$ of the training component set of each $D_{\rm i}$. We use a 10-fold cross-validation on the training data set of each $D_{\rm i}$ to avoid the issue of overfitting. This process divides the training set into 10 groups of equal sample sizes. Training the model on 9 groups, the validation is done on the one remaining group of data points. This happens 10 times such that each data group is made the validation set once.
The average validation TSS from this 10-fold cross-validation is used to decide the model hyperparameters for each data set $D_{\rm i}$. 
We train our models with different values of the model hyperparameters and the optimal values of the hyperparameters $C_{\rm opt}$
are obtained for each $D_{\rm i}$ by maximizing the average validation TSS.
Finally, we choose our operational model hyperparameter ($C_{\rm best}$) by selecting the most frequently appearing $C_{\rm opt}$ among these 20 experimental sets $D_{\rm i}$.
This completes our model optimization process. Now we check the performance of the finalized model and judge the model based on the average test performance over 20 $D_{\rm i}$.

\subsection{ML Models}

\begin{figure*}
    \centering
    \includegraphics[width=.9\linewidth]{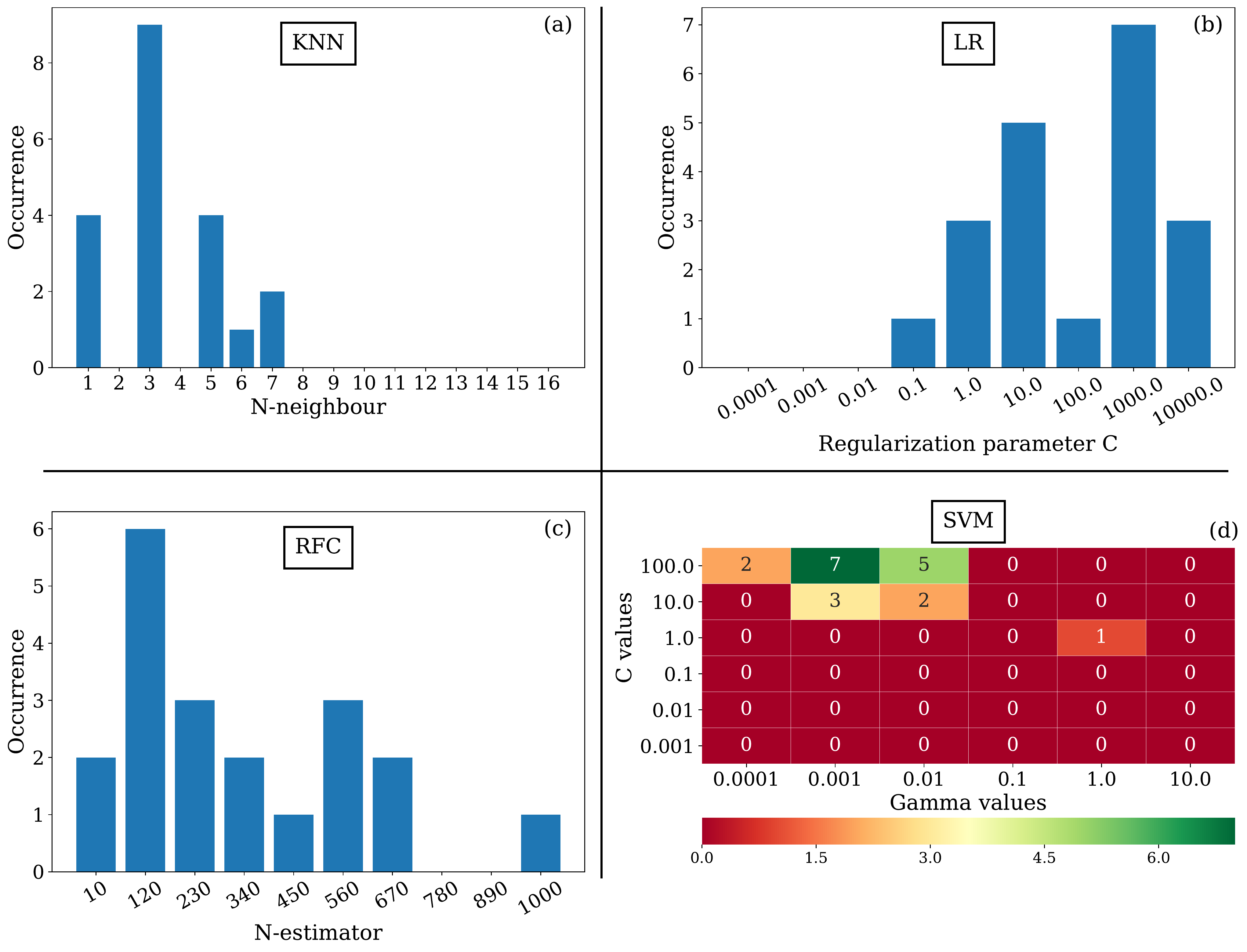}
    \caption{These plots represent the values of the optimal parameters, which give the best TSS score on the validation data set for each experimental data set (D$_{\rm i}$).}
    \label{fig:parameter-search}
\end{figure*}

We use four popular supervised ML algorithms available in the Python scikit-learn library \citep{scikit-learn}.
These four algorithms are K-Nearest Neighbors (KNN), Logistic regression (LR), Random Forest Classifier (RFC) and Support Vector Machine (SVM); they are discussed briefly in this section.
In all four models, the model parameters are tuned properly to get the best achievable performance, which is the TSS in our case. The optimal model parameters of all four models are selected from the histogram plot of Figure \ref{fig:parameter-search}. A general overview of these ML algorithms can be found in \cite{bishop2013pattern} and \cite{MEHTA20191}. 

\begin{enumerate}
\item KNN: This is an instance-based ML
technique that uses instances of training data to compute the machine
classification based on a simple majority vote of ‘k’ number of nearest neighbors of each point \citep{Fix1951}.
When the data set is not large, as in our case, using the KNN classifier poses no disadvantage as it does not create an internal model, which might otherwise use a large memory space. In our model, the weights assigned to each neighbor are equal and the nearest neighbors are calculated using Euclidean distance. The best KNN model is obtained by finding the optimal K, i.e., the optimal value of the number of nearest neighbors to maximize the TSS output. We search for the optimal K value between 1 and 16 in the 20 data sets ($D_{\rm i}$) to get the maximum TSS. Figure \ref{fig:parameter-search}(a) shows the histogram plot of optimal $K$ values for the 20 different data sets.
Since $K=3$ has the highest number of occurrences, it becomes our final choice.
    
    \item LR: This classifier, also known as the log-linear classifier, is a linear classification model that uses the sigmoid function to classify data into discrete categories \citep{MEHTA20191}. This makes it extremely suitable for binary classification problems. Our model uses regularized LR, and is implemented using the LR classifier available in scikit-learn. The only free parameter of this model, which we use, is the regularization parameter $C$, and the most favorable value is estimated from within the range [0.0001, 10000.0], varied with logarithmic increments. In Figure \ref{fig:parameter-search}(b) we can see that the occurrence is at its maximum at $C=1000$; hence we choose 1000 as the optimal $C$ parameter. 
    
    \item RFC: This classifier consists of a large number of individual decision tree classifiers that operate as an ensemble \citep{RF}. Each decision tree is trained on a subset of the entire data set. Generally, decision trees tend to overfit the data and exhibit high variance. Random forests are constructed in such a way so as to decrease the variance. The overall prediction is generated by taking an average of the constituent tree predictions, which tends to cancel out some prediction errors from individual trees. Thus, a large number of uncorrelated trees can produce largely accurate ensemble predictions. Our model uses the RFC available with the scikit-learn package, and the best forest is created by varying the number of trees (also called the 'N-estimator') in the forest from 10 to 1000. We can see from figure \ref{fig:parameter-search}(c) that occurrence is at its maximum for the $N{\text-}estimator=120$.
    Therefore, we select 120 as the optimal value of the number of trees. 

    \item SVM: This is a powerful classification technique \citep{SVM}, and has previously yielded the best results amongst various ML models, when applied to solar flare prediction based on SHARP parameters. SVM works by creating a decision boundary, marked by a subset of training points called support vectors, to separate the positive and negative events in the training data. It uses a kernel function to map the data points to higher-dimensional space. Our model uses a Gaussian radial basis function as the kernel and assigns the class weight in a way that is inversely proportional to the class frequency to handle the class imbalance problem.
    The kernel coefficient gamma ($\gamma$) and the regularization parameter $C$ are varied within the ranges [0.0001, 10.0] and [0.001, 100.0], respectively, to get the best SVM model by comparing the $TSS$ scores. The decision bound From Figure \ref{fig:parameter-search}(d) we can see the optimal values of $C$ and $\gamma$ are 100 and 0.001, respectively, as this combination produces the highest TSS score in 7 out of 20 $D_{\rm i}$. 
\end{enumerate}

\section{Results}

\begin{table*}[]
    \centering
    \begin{tabular}{| C | C | C | C | C | C | C | C | C |}
        \hline
        \multirow{3}{5em}{\centering Classifier Name} & \multicolumn{4}{|c|}{\multirow{2}{20em}{\centering Average Performance Measure in 20 trials when the models are optimized for TSS}} & \multicolumn{4}{|c|}{\multirow{2}{12em}{\centering An Example of Confusion Matrix Elements}} \\ 
        
        \multirow{1}{5em}{\centering } & \multicolumn{4}{|c|}{\multirow{1}{10em}{\centering}} & \multicolumn{4}{|c|}{\multirow{1}{12em}{\centering}} \\         
        \cline{2-9}
        & SSW & CSW & TSS & MAC & TN & TP & FN & FP \\
        \hline
        \hline
        KNN & 0.887 $\pm$ 0.040 & 0.990 $\pm$ 0.006 & 0.938 $\pm$ 0.019 & 0.969 $\pm$ 0.010 & 663 & 110 & 6 & 3 \\
        \hline
        \multirow{2}{5em}{\centering Random forest} & 0.898 $\pm$ 0.042 & 0.989 $\pm$ 0.008 & 0.944 $\pm$ 0.020 & 0.972 $\pm$ 0.010 & 664 & 101 & 6 & 2 \\
        & & & & & & & &\\
        \hline
        \multirow{2}{5em}{\centering Logistic regression} & 0.959 $\pm$ 0.033 & 0.975 $\pm$ 0.009 & 0.967 $\pm$ 0.018 & 0.983 $\pm$ 0.009 & 669 & 94 & 1 & 9 \\
        & & & & & & & &\\
        \hline
        SVM & 0.956 $\pm$ 0.031 & 0.974 $\pm$ 0.010 & 0.965 $\pm$ 0.017 & 0.983 $\pm$ 0.009 & 656 & 105 & 2 & 10 \\
        \hline
    \end{tabular}
    \caption{Performance of classifiers trained with the best hyperparameters deduced via Grid Search and 10-fold cross-validation over 20 randomly shuffled data sets. The confusion matrix elements correspond to a test data set in $D_{\rm i}$ whose TSS is closest to the determined mean.}
    \label{tab:classif_score}
\end{table*}

\begin{figure}
    \centering
    \includegraphics[width=\linewidth]{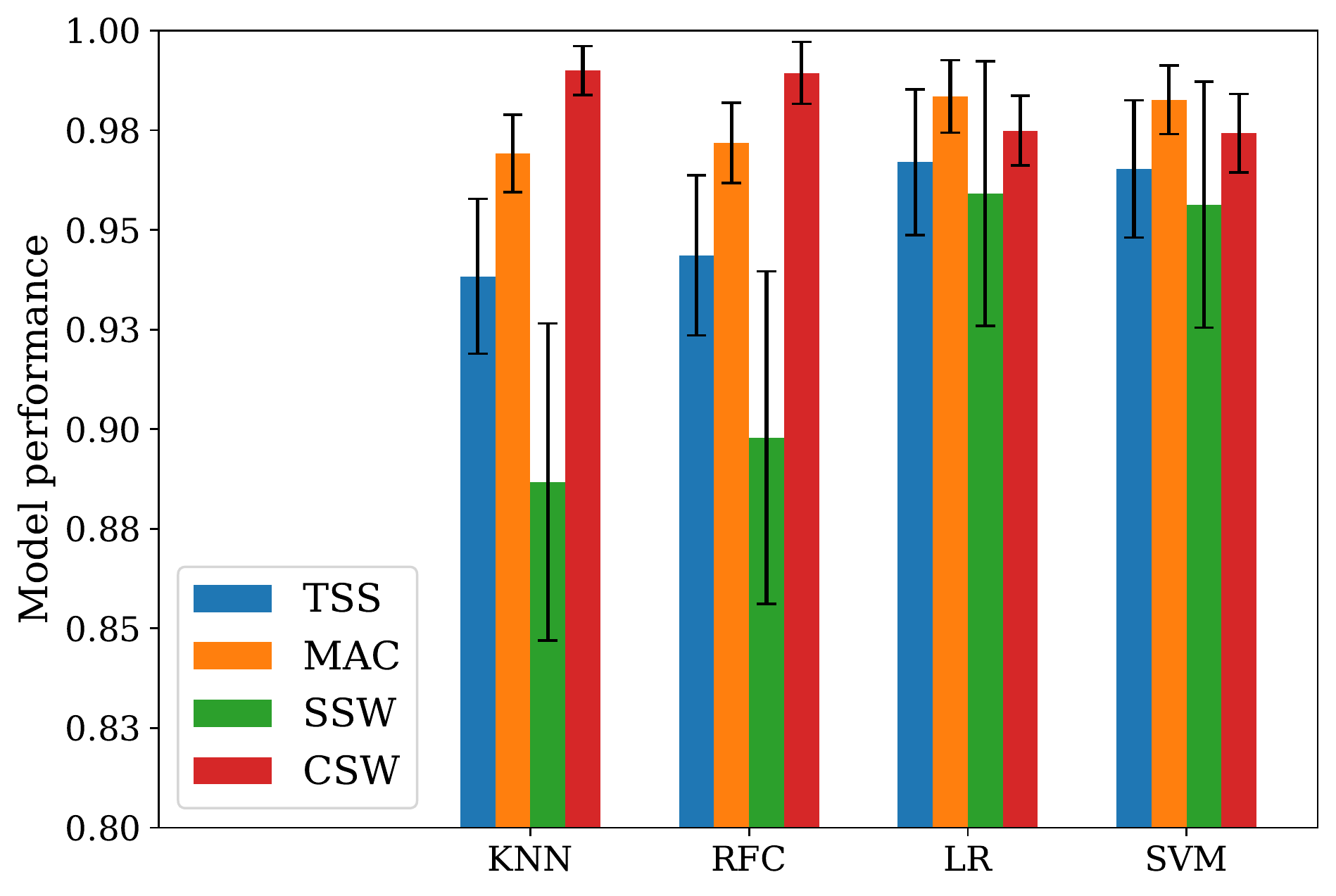}
    \caption{This plot depicts the classification performance of all four models.}
    \label{fig: model performances}
\end{figure}

Each ML classifier is trained with the finalized hyperparameters ($C_{\rm best}$) on the training set of each $D_{\rm i}$, and then the trained model is applied on the test set in that $D_{\rm i}$. The average and standard deviation of the performance metrics over these 20 test sets are reported in Table \ref{tab:classif_score}. We find that all of these models work reasonably well for identifying flaring and nonflaring ARs. The performance of both LR and SVM is very similar, better than that of KNN and RFC. The average TSSs of LR and SVM are 0.967 and 0.965, respectively. Therefore, we claim that LR and SVM are equally good in terms of performance. For further analysis we primarily focus on LR because of its marginally higher TSS value.
The comparison of the four ML classifiers is depicted in Figure \ref{fig: model performances}. We achieve a remarkable MAC of 0.983 for both LR and SVM.
The SSW is also much higher for LR (and SVM) than for RFC or KNN, which indicates the suitability of LR (and SVM) in highly active space weather circumstances. Also the close values of SSW and CSW tells us of the unbiased nature of the model predictions. Clearly, LR/SVM is a better choice over KNN/RFC for having a very similar SSW and CSW scores.  
An example of the confusion matrix elements corresponding to a seed value of $D_{\rm i}$ with the TSS close to the mean value is also shown in Table \ref{tab:classif_score}.

\begin{figure}
    \centering
    \includegraphics[width=\linewidth]{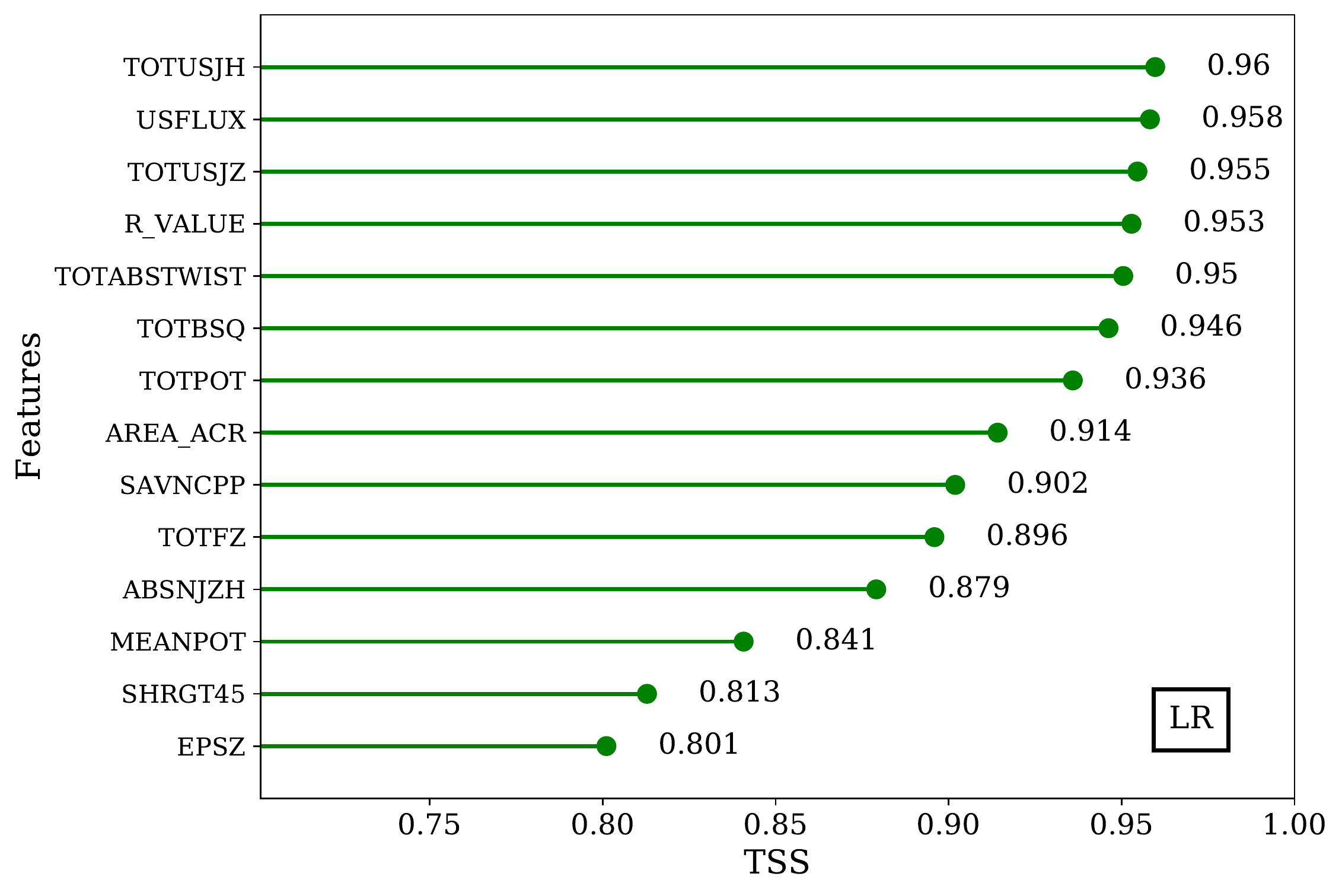}
    \caption{Average TSS scores of individual features for LR. Each score is obtained by training the LR model with a single parameter as input data, averaged over the output of the 20 experimental data sets.}
    \label{fig:individual_tss_LR}
\end{figure}

To understand which AR parameters are more useful in determining the flaring capability of an AR, we train our models with the 14 AR parameters individually. The outcome of this experiment for LR is presented in Figure \ref{fig:individual_tss_LR}, where all these AR parameters are plotted along the y-axis in ascending order of their individual TSSs.
This implies that the topmost parameter in the y-axis is the most significant one having the highest individual classification capability and as we move downward, we find parameters of lesser importance.

\begin{figure}
    \centering
    \includegraphics[width=\linewidth]{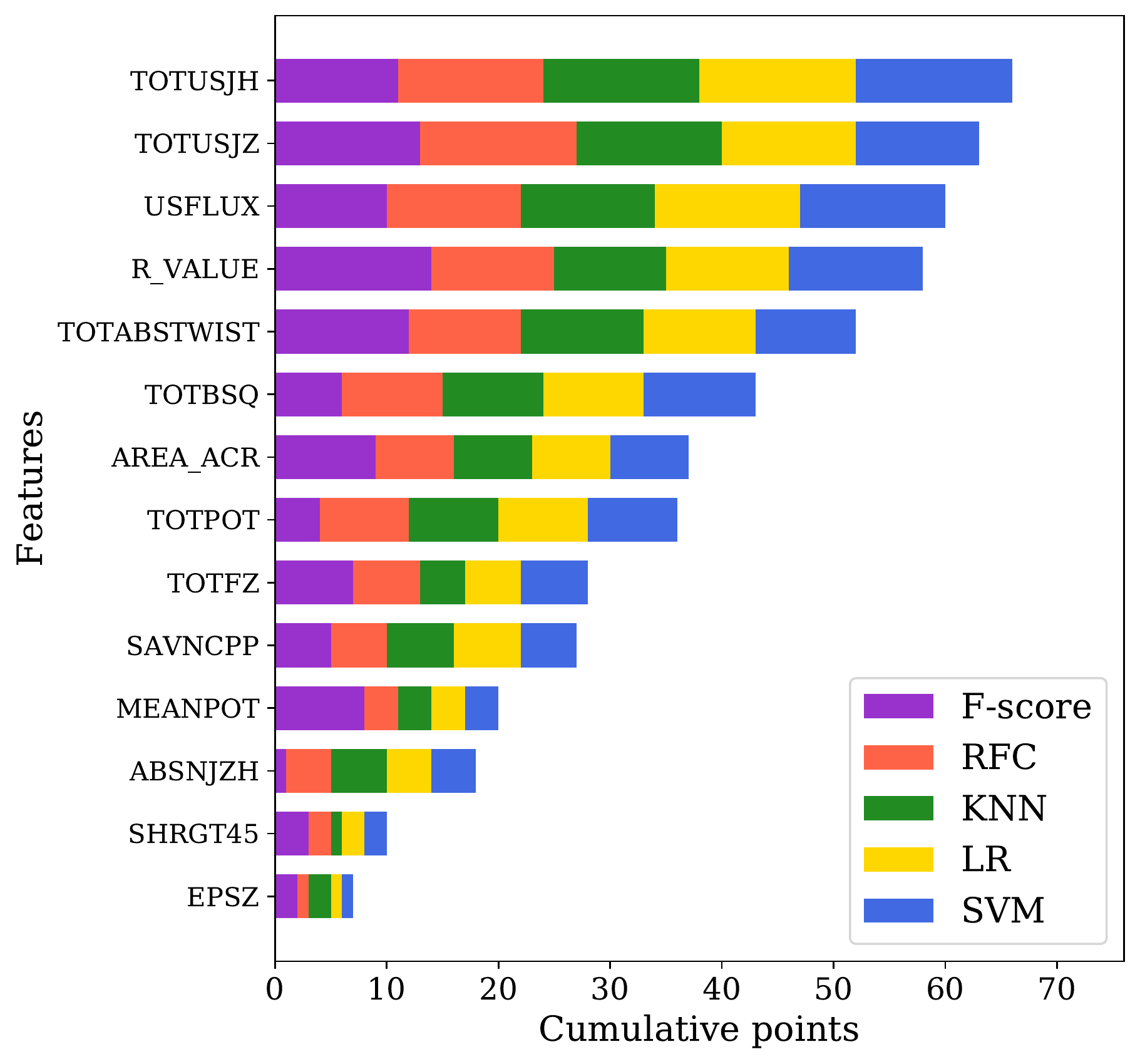}
    \caption{Feature ranking based on cumulative points obtained from F-statistics, RFC, KNN, LR, and SVM. The top-scoring feature in each model gets 14 points, while lowest-scoring feature gets one point. The points are added for each feature and then the features are ranked accordingly.}
    \label{fig:cumulative ranking}
\end{figure}

The ranking of input features based on the individual TSSs depends on the ML model used, and can moderately differ for different models. For a particular ML algorithm, feature ranking may also depend on the model hyperparameters.
Hence to get a more general global ranking of features, we follow a marking scheme in which we assign points (ranging from 1 to 14) to each parameter based on its individual TSS ranking for each of the four models and the univariate F-score ranking.
For example, for LR 14 points are assigned to TOTUSJH for its highest TSS, whereas EPSZ gets 1 point based on Figure \ref{fig:individual_tss_LR}. Finally we add up all the points for each feature from the different models to get a cumulative ranking as shown in Figure \ref{fig:cumulative ranking}. 

\begin{figure*}
\gridline{\fig{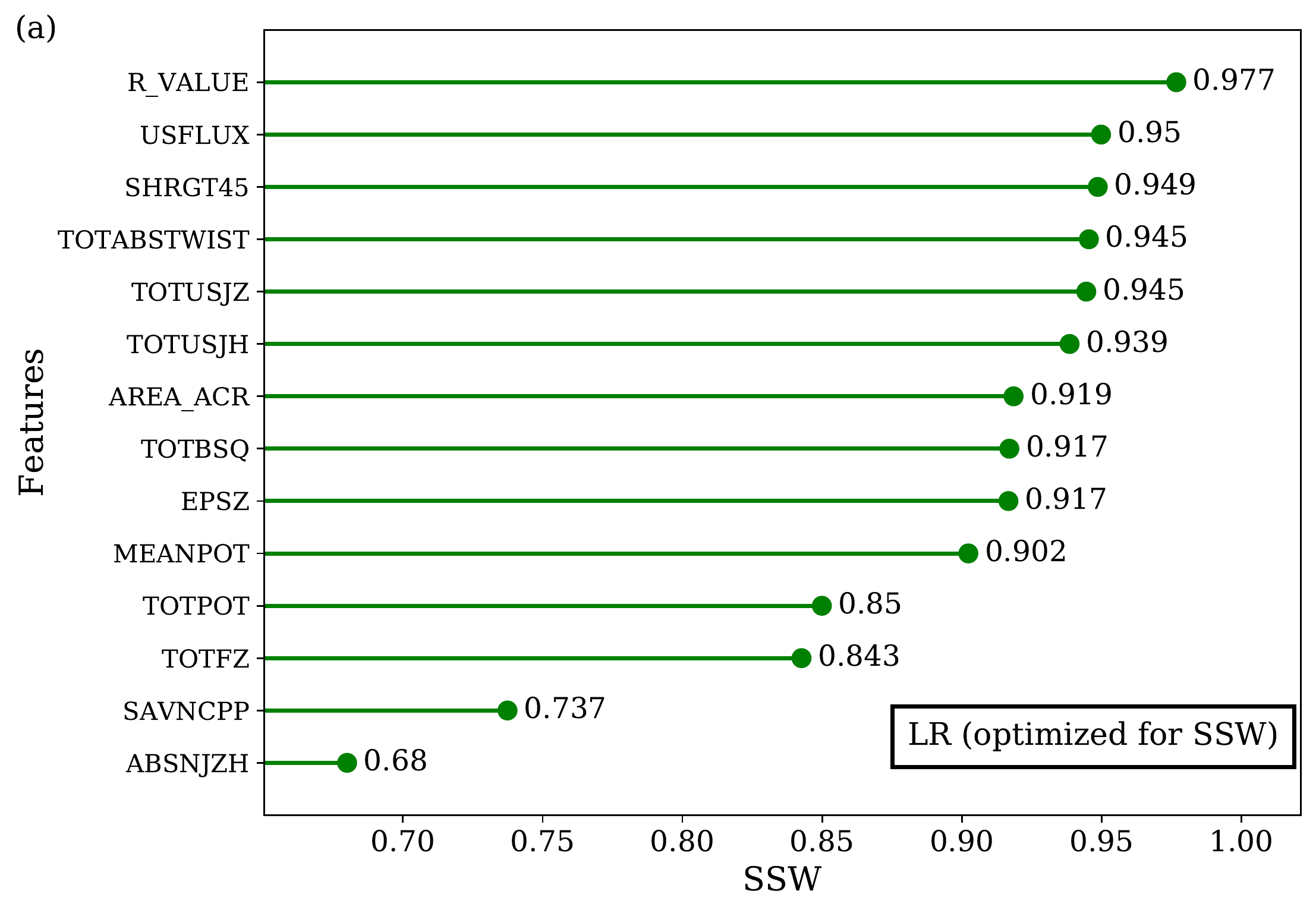}{0.48\textwidth}{}
          \fig{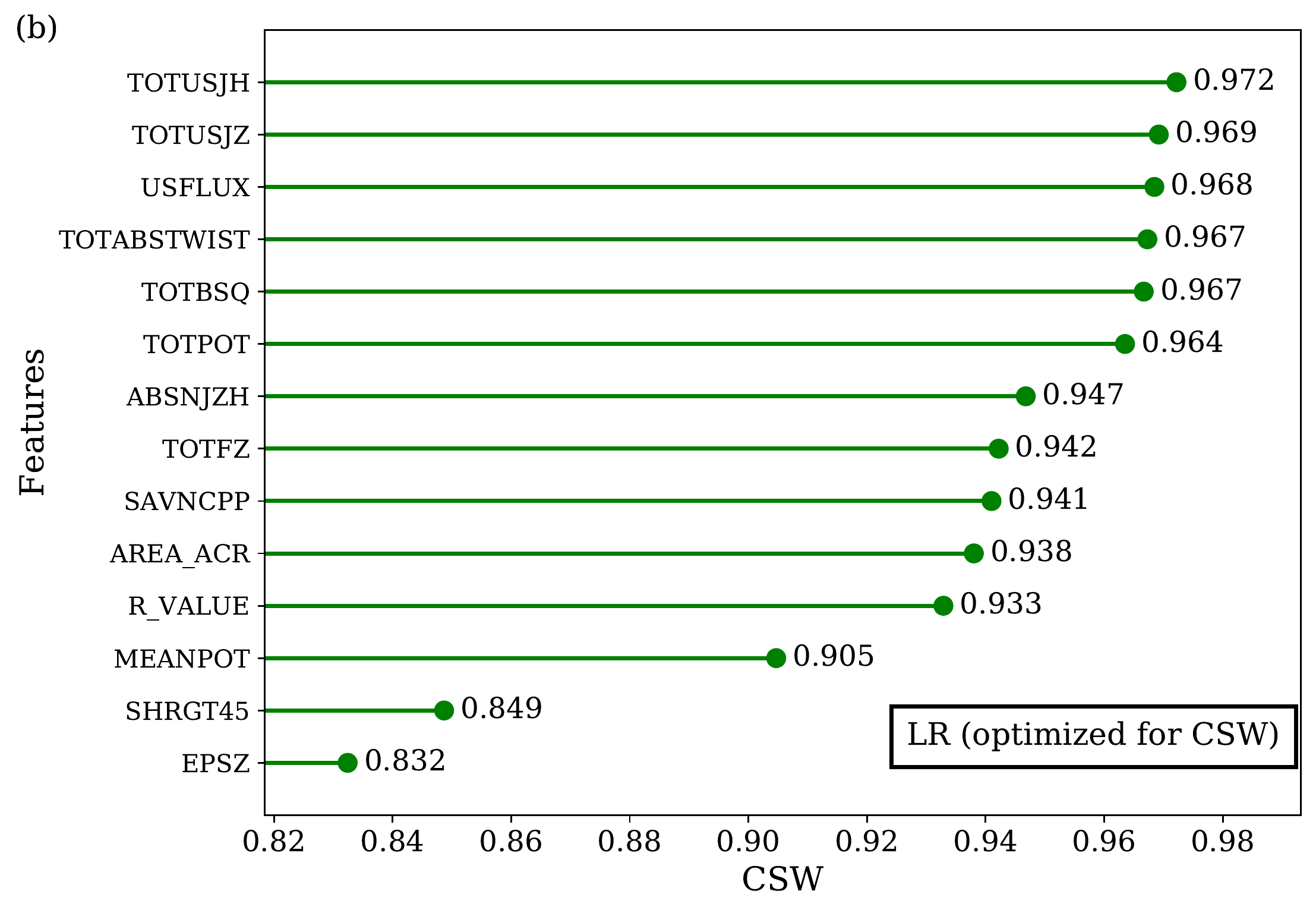}{0.48\textwidth}{}
          }
\caption{The left panel (a) shows the feature ranking for the LR model when the model is optimized for the metric SSW and the panel (b) on the right shows the feature ranking when the LR model is optimized for the CSW metric.}
\label{fig:individual ranking for ssw and csw}
\end{figure*}

We further optimize the LR model by tuning the model hyperparameter for maximizing the SSW and CSW metrics to see how the ranking of input magnetic features changes for these two newly introduced space weather metrics. The left panel in Figure \ref{fig:individual ranking for ssw and csw} shows the ranking of input features according to their individual SSW scores and the right panel shows the feature ranking with respect to the CSW score. We can see that the ranking of R\_VALUE, SHRGT45, and EPSZ goes down significantly on optimizing our LR model with CSW instead of SSW. Both R\_VALUE and SHRGT45 shift downward in ranking by 10 due to this change in performance metric. On the other hand, features like ABSNJZH, SAVNCPP, TOTPOT, and TOTFZ perform better on CSW, reflecting an upward shift in the feature ranking (Figure \ref{fig:individual ranking for ssw and csw}.)

\begin{figure}
\centering
\includegraphics[width = \linewidth]{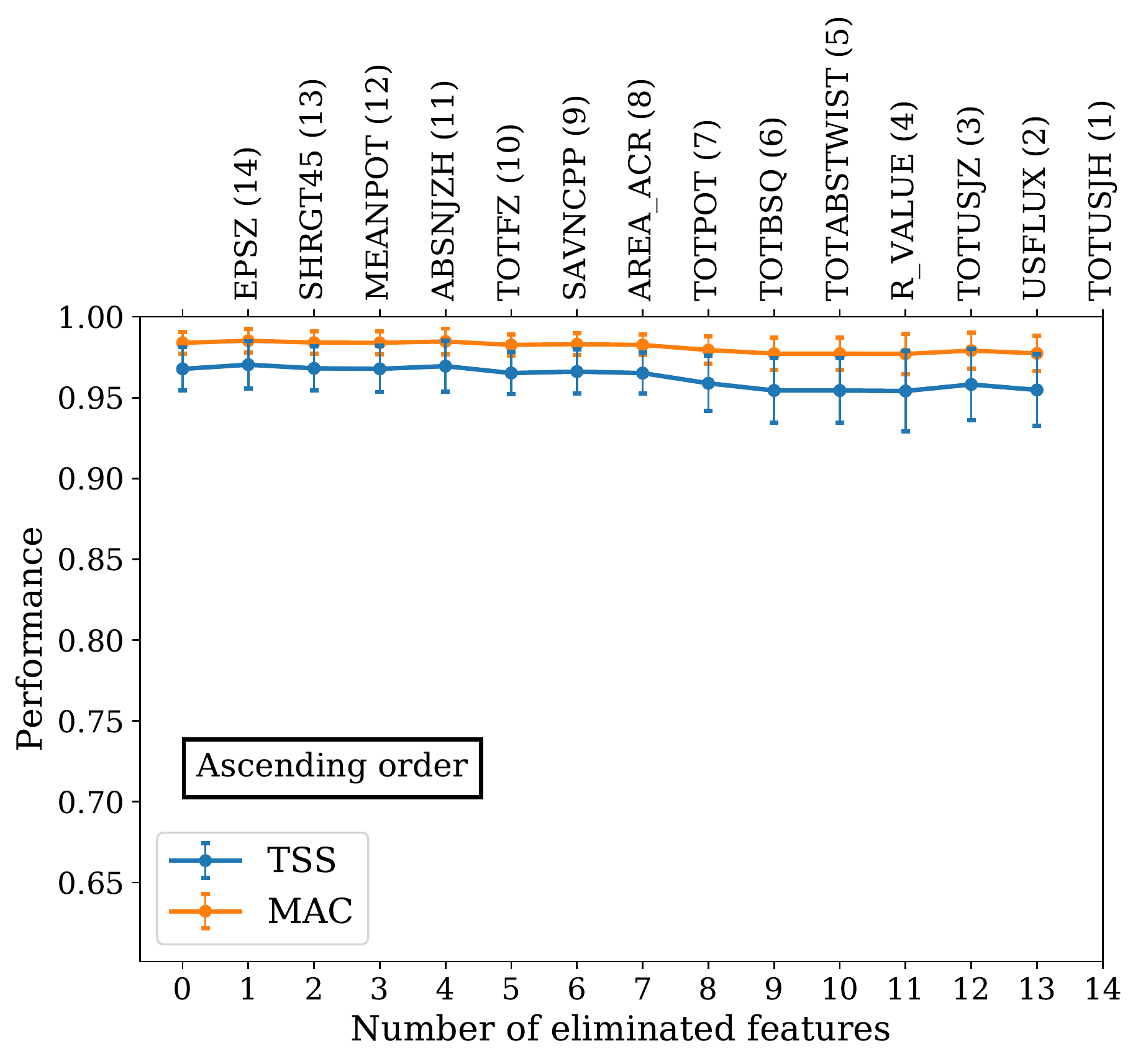}
\includegraphics[width = \linewidth]{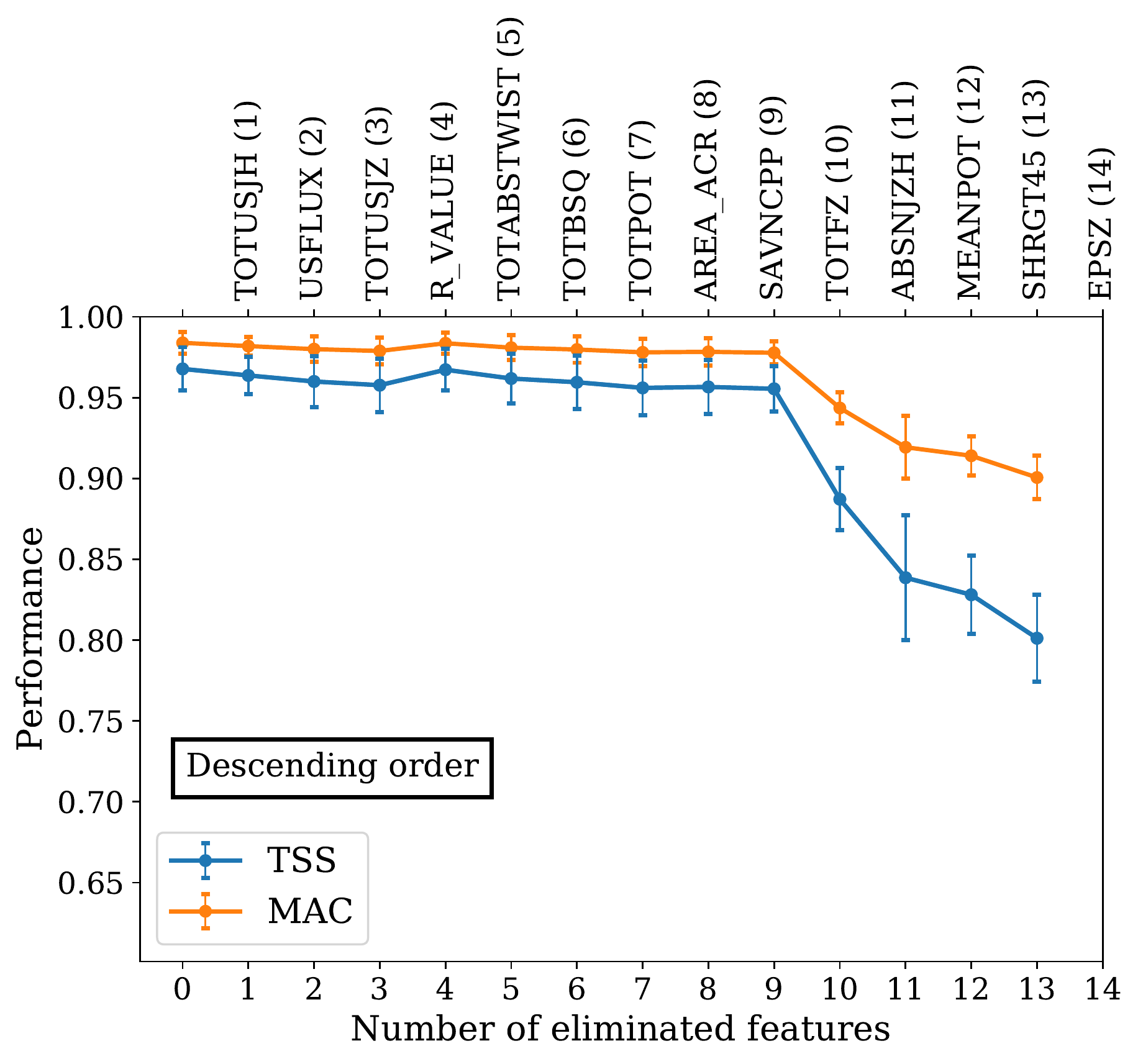}
\caption{Variation of LR performance with number of input features. The experiment is performed over all 20 experimental data sets ($D_{\rm i}$). In each step, the eliminated feature along with its rank is indicated on the top x-axis whereas the bottom x-axis indicates the total number of eliminated features at that step.}
\label{fig:tss variation}
\end{figure}

\begin{figure*}
    \centering
    \includegraphics[width=\linewidth]{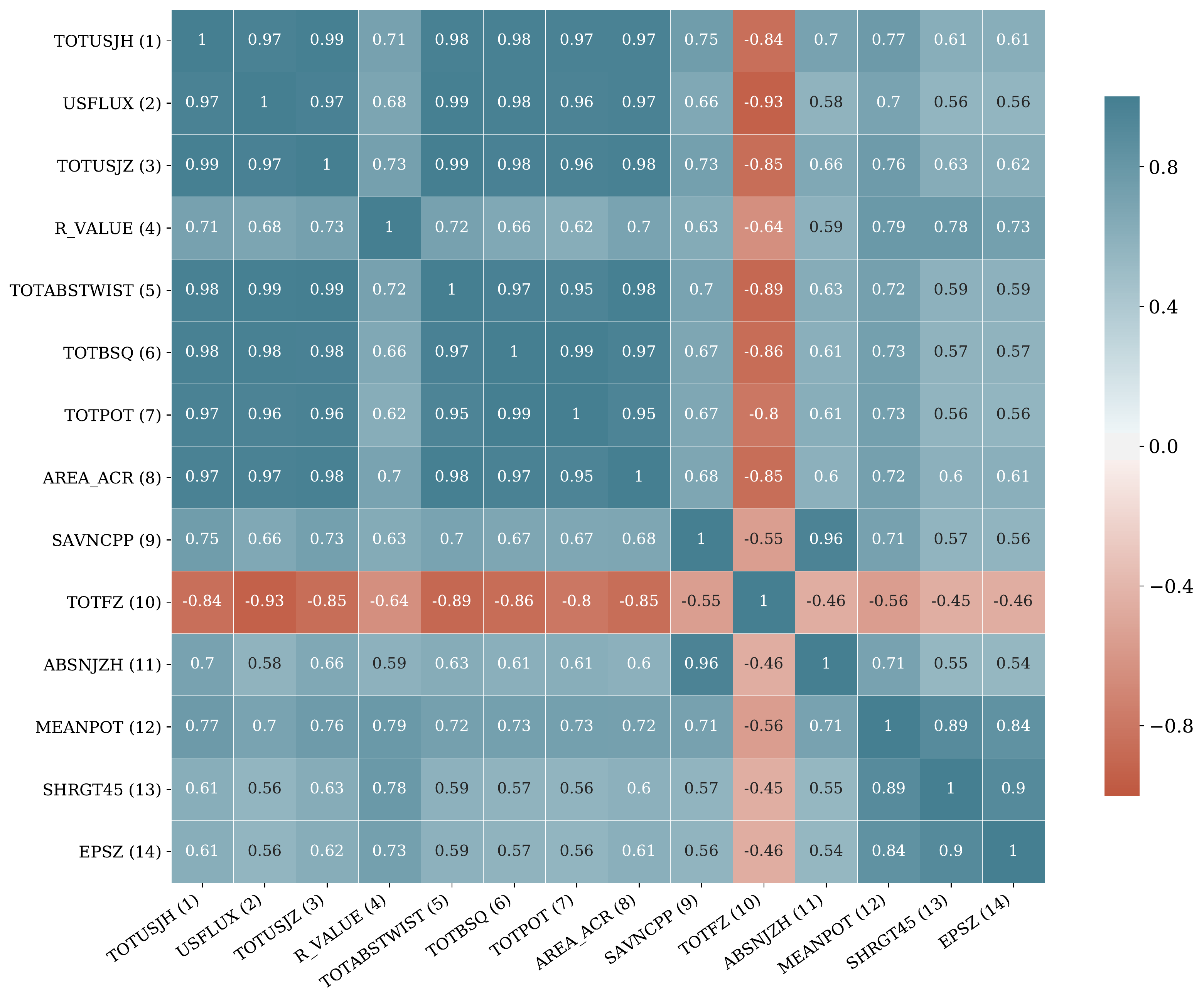}
    \caption{Correlation matrix for the input features, which are arranged in the order of their individual TSS ranking with LR. The ranks are shown in brackets next to the feature keywords.}
    \label{fig: correlation matrix}
\end{figure*}

To study the dependence of model performance on the number of input AR features, we train our LR model by eliminating input features one by one and check the model performance at each step. The elimination is done by following both the ascending and descending order of ranking based on the individual TSS, the result of which is represented in Figure \ref{fig:tss variation}.  We can see that when we eliminate the features in ascending order of their ranking the model performance does not change much. This is expected because the more important features are eliminated at the last steps. On the other hand, for the descending order we see a drastic fall in model performance when the number of eliminated features increases beyond 9. This indicates a significant loss of correlation with the output labels at each step beyond this point. The plateau in the descending-order plot of Figure \ref{fig:tss variation} is only possible if the top-ranked features are highly correlated among themselves, causing no significant loss of information when these features are thrown out. A study by \cite{Hazra2015} also confirms the correlation amongst integrated magnetic features, showing a connection between AR magnetic properties and coronal X-ray flux. The correlogram presented in Figure \ref{fig: correlation matrix} confirms this, with all top eight features, excluding R\_VALUE, being highly intercorrelated.
One possible reason behind this high correlation could be that they are extrinsic features, or in other words, that their values depend on the size of the AR as they represent the sums of physical quantities over the entire AR. As correlated features do not provide new information, we group features with correlation constants $> 0.9$ and train our model by picking up the top-performing feature from each group. Following this scheme we select six features: TOTUSJH, R\_VALUE, TOTFZ, SAVNCPP, MEANPOT, and SHRGT45. When trained with these features only, the LR classifier gives an average TSS and MAC values of 0.962 and 0.981, respectively (with the SSW and CSW of 0.956 and 0.968), which are close to our primary model performance with 14 features.

\section{Conclusions}

With the advancement of new technologies, especially in satellite-based telecommunications and navigational networks, a significant fraction of our technological assets have become increasingly vulnerable to space weather disturbances. This has resulted in growing demand for reliable space weather forecasts.
Solar flares strongly influence the space weather, which is why we address the problem of predicting solar flares using their source region characteristics. In this work, we have built a high performance operational LR classifier that can differentiate solar ARs based on their flaring capabilities.
We have compared four supervised ML models, all of which perform quite well in classifying ARs into positive/flaring and negative/nonflaring categories. The method we follow is statistically unbiased due to the use of 20 randomly shuffled replicas of the primary data set for measuring model performance.
The LR classifier delivers the highest average TSS score of 0.967 $\pm$ 0.018 closely followed in performance by the SVM classifier. 

While a direct comparison of model performance between our algorithms with that in previous studies may not be appropriate due to subtle differences in the data selection scheme and the size of the database used, we do note that in the context of the TSS, we achieve a higher performance score relative to earlier classification attempts with supervised ML algorithms (e.g., \citealt{Bobra2015, Nishizuka2017, Florios2018}).

\begin{figure*}
    \centering
    \includegraphics[width=.85\linewidth]{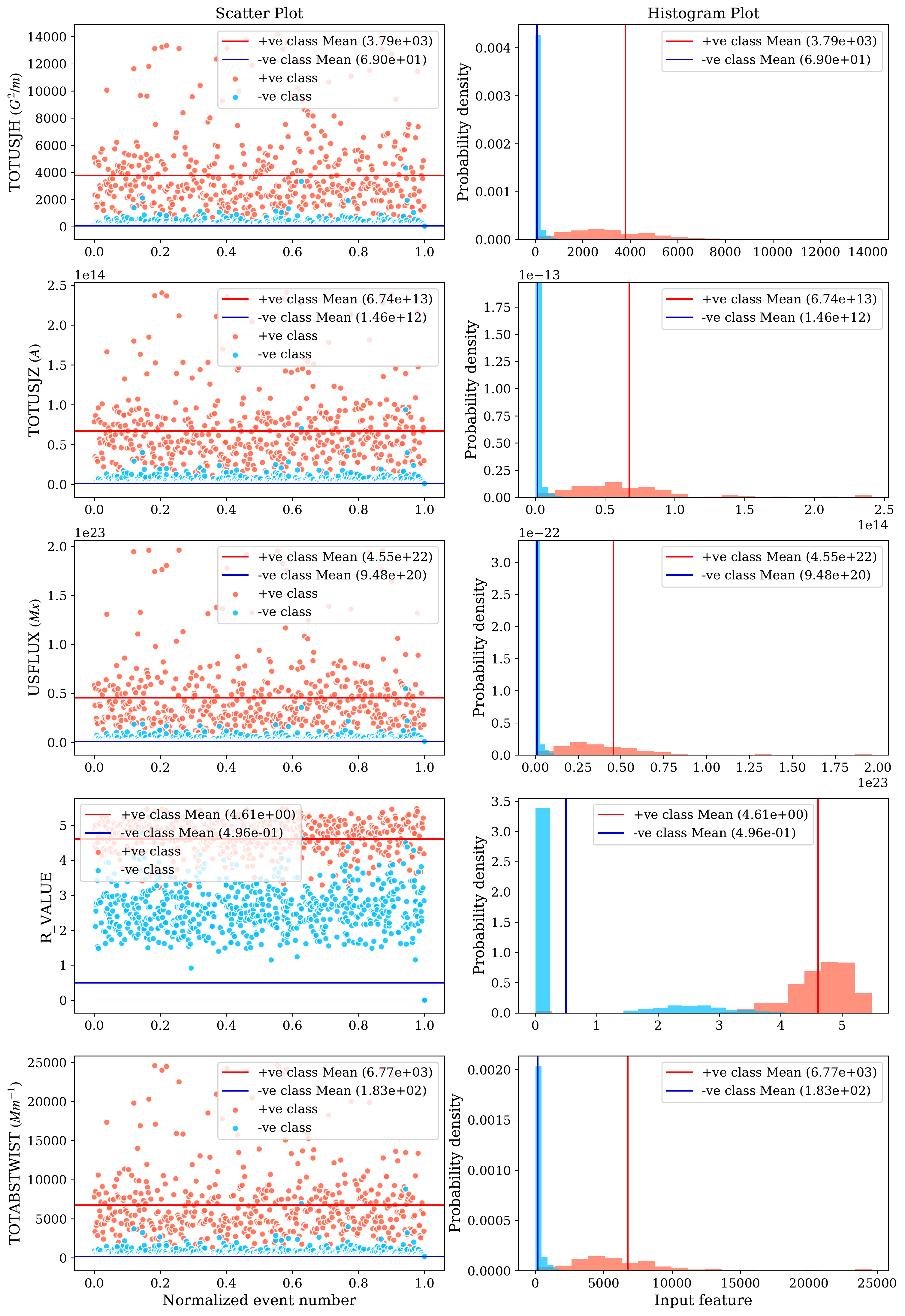}
    \caption{Distribution of magnetic features in positive and negative classes. The left column shows scatter plots of the top five input features according to the cumulative feature ranking. The $X$-axis of the scatter plots is the normalized event number, which is the number of events divided by the total number of events in that class. The mean values for both the classes are shown. In the right column, histogram plots of the probability density are shown for the corresponding features.}
    \label{fig: feature distribution}
\end{figure*}

One possible reason for our achieving a high TSS could be the exclusion of C-class events in the data preparation stage. The distribution of the top five input features in our data set is shown in figure \ref{fig: feature distribution}, where we can see a clear separation between two clusters of data points for the two different classes. This ensures that our data set is easily separable with two distinct classes in feature space. Other possible reasons could be the different event selection scheme and the larger temporal coverage of our data set and also it is important to note that each entry in the negative class comes from a different SHARP region ensuring no repetition of AR patches in the nonflaring class.

In addition to achieving a high TSS, we find that a global indicator of magnetic twist, estimated by the feature TOTABSTWIST, plays an important role in predicting AR flare potential. Although TOTABSTWIST comes in the fifth position of the cumulative feature ranking, other twist-related parameters including VTWIST and MEANALP, are not found to play a significant role.  

We have also introduced two new performance indicators, termed SSW and CSW, which are useful in comparing model performance depending on the operational space weather condition one wishes to lay more emphasis on. For example, when the solar activity is high, we may wish to get a reliable all-clear forecast for executing specific time-critical tasks that are susceptible to space weather. So, depending on the application and operational space weather scenario,
SSW and CSW can provide more meaningful operational intelligence than the TSS alone. We can also get an estimate of the model bias toward a specific forecast by examining the difference between SSW and CSW. With these two indicators, we see that KNN and RFC are more biased toward the negative class (as CSW is much higher than SSW) than SVM and LR.
Because of the larger size of the negative class, a classifier's forecast may become biased toward it. But our analysis shows that classifiers such as LR and SVM can be suitably optimized to minimize the class imbalance problem significantly.

We have also studied the relative importance of input features in terms of their ability to classify the flaring and nonflaring ARs. Based on the global ranking of Figure \ref{fig:cumulative ranking}, we have identified key magnetic features that are responsible for the flare potential of an AR. The total unsigned current helicity, the total unsigned vertical current, the total unsigned magnetic flux, the flux near strong-field high-gradient neutral line, and the total absolute twist are the major deciding factors for AR flare potential. It is important to note that all of the highly ranked features in Figure \ref{fig:cumulative ranking} denote extensive or net properties of an AR, except for R\_VALUE. This reaffirms previous findings \citep{Welsch2009, Hazra2020} that extensive parameters contribute more to forecasting algorithms than intensive parameters. The only nonextensive feature that performs well is R\_VALUE, indicating that it contains some unique information regarding flaring potential.

Our analysis shows that for a given classifier, the ranking of input magnetic features differs based on the choice of the performance metric.
For example, the ranking of R\_VALUE and SHRGT45 drops from 1st to 11th and from 3rd to 13th, respectively, when we switch to CSW from SSW as the model-optimizing metric. The reason behind this downward shift in feature ranking can be explained by examining the feature distribution plot. We have investigated this trend further by analyzing the feature distribution and found well-separated peaks in the frequency distribution for binary classes; we believe this might be the reason for these features perform well in discriminating between the classes. For the nonflaring class, these distributions are sparse and skewed towards the central peak of the flaring class. This reduces the ability of perfect identification of nonflaring events, especially those distributed near the flaring-class mean. As the CSW determination involves only the nonflaring class, the ranking of these features that are not well separated from the mean of the other class shifts down. In Figure \ref{fig: feature distribution} (fourth row, right panel) we can see that R\_VALUE shows a bimodal distribution leading to a further increase in variance for the nonflaring class. This bimodal distribution of R\_VALUE is due to the absence of strong-gradient magnetic PILs in many nonflaring ARs. On the other hand, features whose ranking improves on optimizing with CSW have in general a sharp peak in the distribution for the nonflaring class with very small variance. This helps these features (ABSNJZH, SVANCPP, TOTFZ, and TOTPOT) to correctly identify the nonflaring events.

We also find that the model performance has very low dependency on the number of input features especially when the input features are highly correlated. We have shown that a high model performance could be maintained even with a smaller set of input magnetic features, selected carefully to reduce internal correlation. Our work brings to the fore key properties of parameter-based ML flare forecasting that can be utilized in future works to develop more robust flare forecasting models. Finally, we anticipate our comprehensive analysis will lead to operational flare forecasting with higher efficiency and higher precision. 

\section*{Acknowledgements}

The Center of Excellence in Space Sciences India at IISER Kolkata is funded by the Ministry of Education, Government of India. The authors acknowledge funding from the Ministry of Education SPARC Project grant SPARC/2018-2019/P746/SL to IISER Kolkata, Nordita/Stockholm University, the KTH Royal Institute of Technology, and IIT Kharagpur. S.S.\ thanks the University Grants Commission, Government of India, for a senior research fellowship. O.G. thanks the Department of Science and Technology, Government of India, for an INSPIRE scholarship. D.N.\ thanks the Wenner Gren Foundation for a visiting professorship at Nordita, Stockholm University. Nordita is sponsored by NordForsk. All the magnetic field data are either collected or calculated from SDO/HMI SHARP data, maintained by the Joint Science Operation Center. The authors have utilized the drms open-source software package \citep{Glogowski2019} to access data from HMI. The authors acknowledge usage of data from the XRT Flare Catalogue and GOES flare database to acquire the information on flaring events and their corresponding ARs. All ML algorithms have been implemented using the scikit-learn package \citep{scikit-learn} on Python.

\vspace{2mm}\noindent {\large\em Data Availability:} The input magnetic feature data set used for this study is freely available on Zenodo: \url{https://doi.org/10.5281/zenodo.5498347}.

\bibliographystyle{aasjournal}
\bibliography{ref}

\end{document}